# On the weighting field and admittance of irradiated Si-sensors

R. Klanner and J. Schwandt

*Abstract*—In this paper the weighting field $E_W$ and the frequency dependence of the admittance Y of $n^+p$ pad sensors irradiated by 24 GeV/c protons to equivalent fluences in the range $\Phi_{eq} = 3$ to $13 \times 10^{15}$ cm$^{-2}$ are investigated. 1-D TCAD simulations are used to calculate $E_W$. For $\Phi_{eq} \lesssim 10^{13}$ cm$^{-2}$ $E_W$ depends on position and time. However, for higher $\Phi_{eq}$ the time constant $\tau$ is much longer than the typical electronics readout time and $E_W = 1/d$ (d = sensor thickness). It is demonstrated that the increase of the resistivity of the Si bulk with irradiation is responsible for the increase of $\tau$.

The admittance Y of irradiated pad sensors has been measured for frequencies between $f = 100$ Hz and 1 MHz and voltages between 1 and 1000 V at –20 °C and –30 °C. For $f \lesssim 1$ kHz the parallel capacitance $C_p$ shows a $f$ dependence. A model with a position-dependent resistivity is able to describe the data.

It is concluded: 1. The weighting field of a highly irradiated sensor is the same as the weighting field of a fully depleted sensor before irradiation. 2. Models with a position-dependent resistivity describe the frequency dependence of $C_p$ for irradiated sensors.

## I. Introduction

This contribution addresses two questions: 1. "What is the weighting field $E_W$ of radiation-damaged silicon sensors?"; 2. "What is the reason of the frequency dependence of the capacitance of radiation-damaged silicon sensors?" Common answers are: 1. "$E_W$ is time independent and equal to the $E_W$ of the fully depleted sensor before irradiation"; 2. "The frequency response of the radiation-induced traps." To study these questions planar pad sensors are studied for which 1-D simulations are valid. For 1. an analytical model for a partially depleted non-irradiated sensor is fitted to the results of TCAD simulations of irradiated sensors. For 2. the admittance as a function of frequency and voltage for sensors irradiated by 24 GeV/c protons to 1 MeV neutron equivalent fluences of 3, 6, 8 and $13 \times 10^{15}$ cm$^{-2}$ has been measured, the data fitted by an electrical model and the model parameters determined. The hardness factor used is 0.62.

## II. Comparison of a non-depleted sensor before irradiation with an irradiated sensor

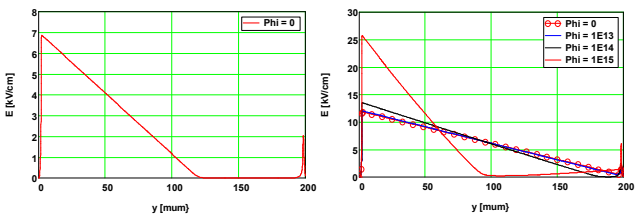

Fig. 1. Simulated electric field $E$ in a 200 μm thick $n^+p$ silicon pad sensor with a B-doping of $3.8 \times 10^{12}$ cm$^{-3}$. Left: Before irradiation at 40 V. Right: After irradiation to fluences between 0 and $10^{15}$ cm$^{-2}$ at 120 V. The depletion voltage before irradiation is $\approx 70$ V. The $y$ direction is normal the sensor surface and the $n^+p$ junction is at $y = 0$.

R. Klanner and J. Schwandt are with the University of Hamburg, Germany (email: Robert.klanner@desy.de and Joern.Schwandt@desy.de).

Fig. 1 compares the simulated electric field in a 200 μm thick $n^+p$ pad sensor at 40 V before irradiation to the electric field at 120 V after irradiation to different 1 MeV equivalent neutron fluences $\Phi_{eq}$. For $\Phi_{eq} = 10^{15}$ cm$^{-2}$ the fields for the irradiated and non-irradiated sensor are similar: A linear decrease from the $n^+p$ junction followed by a low-field region and a narrow high-field region at the $pp^+$ junction. This similarity suggests that a model for a non-depleted sensor before irradiation could be helpful for the understanding of an irradiated sensor at significantly higher voltages.

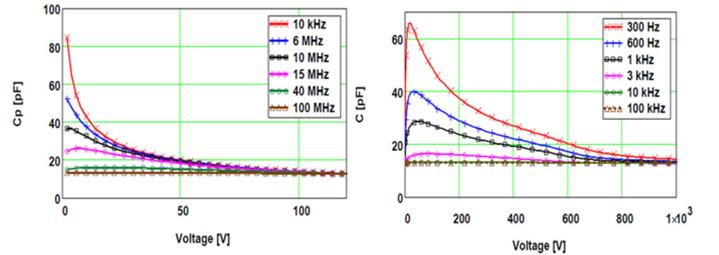

Fig. 2. Left: Simulated voltage dependence of the parallel capacitance $C_p$ of a partially depleted sensor for different frequencies before irradiation. Right: Measured voltage dependence of $C_p$ for the same sensor after irradiation to $\Phi_{eq} = 6 \times 10^{15}$ cm$^{-2}$. Note the differences in the frequency values.

Fig. 2 compares the calculated voltage dependence of the parallel capacitance $C_p$ of a partially depleted non-irradiated sensor to the measured $C_p$ after irradiation to $\Phi_{eq} = 6 \times 10^{15}$ cm$^{-2}$ for different frequencies. The data have been taken at –30 °C. The dependencies are similar. However the frequency at which $C_p$ changes differs by about 4 orders of magnitude, which is similar to $\rho_{intr}/\rho \approx 3 \times 10^4$, the ratio of the intrinsic resistivity at –30 °C to the resistivity of the non-irradiated Si. Thus one might expect that the resistivity in the low-field region of irradiated sensors causes the frequency dependence of $C_p$.

## III. Weighting Field

The weighting field $\vec{E}_W(\vec{r})$ describes the capacitive coupling of a charge at position $\vec{r}$ to the readout electrode. It has been introduced to calculate the induced current in vacuum tubes [1, 2]. The charge induced by a charge $Q_0$ moving from $\vec{r}_1$ to $\vec{r}_2$ is $Q_0 \cdot \int_{\vec{r}_1}^{\vec{r}_2} \vec{E}_W(\vec{r})\, d\vec{r}$. For a pad sensor, which can be treated as a 1-D problem, $\vec{E}_W(\vec{r})$ has only a $y$ component which depends only on the depth $y$. For a fully depleted sensor of thickness $d$, $E_W(y) = 1/d$. If there are free charge carriers in the sensor, which is the case for a partially depleted sensor, a time-dependent $E_W$ is required [3]. The reason is that a charge moving at $t = 0$ from $y$ to $y + \Delta y$ produces an electric field and it takes a finite time until the equilibrium state is reached again. For a partially depleted pad sensor with a depletion depth $w$, the time-dependent weighting field for $t > 0$ is [4, 5]

$$E_W(y,t) = \begin{cases} 1/w - (1/w - 1/d) \cdot e^{-t/\tau}, & 0 < y \leq w \\ 1/d \cdot e^{-t/\tau}, & w < y < d \end{cases} \quad (1)$$

with $\tau = \varepsilon_{Si} \cdot \rho \cdot d/w$ and $\rho$ the resistivity of the non-depleted region. For $t \to 0$ $E_W = 1/d$ in the entire sensor as the free charge carriers had no time to move. For $t \to \infty$ in the depleted region $E_W \to 1/w$, the value for a sensor of depth $w$, whereas in the non-depleted region $E_W \to 0$. In the derivation of Eq. 1, a sharp transition between the depleted and the non-depleted region has been assumed and the $n^+$ and $p^+$ implantations at $y = 0$ and $y = d$ ignored. This is not the case for the simulation with SYNOPSYS TCAD, where Ramo's theorem [2] with a 50 ps voltage ramp by 1 V at the readout electrode has been used. Fig. 3 compares $E_W(y)$ of a pad sensor for $w \approx 80$ μm the TCAD simulation to the analytical calculation using Eq. 1 for different times $t$ after the voltage ramp. Except for the regions around $y = 0$, $d$ and $w$, the calculations agree.

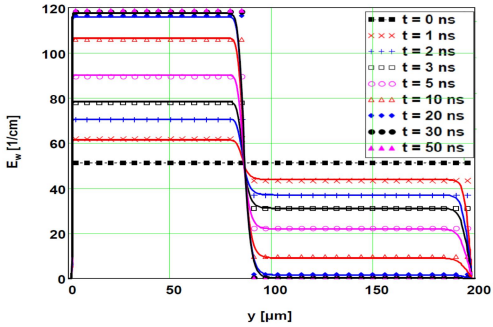

Fig, 3. $y$-dependence of $E_W$ of the non-irradiated sensor at 20 V (depletion depth $w$ = 80 μm) for different times $t$. The symbols are the results of the analytical and the lines of the TCAD calculation.

The TCAD simulation of the irradiated sensor uses the model HPTM [6] with 2 acceptors and 3 donor traps and parameters tuned to the measurement results of pad diodes irradiated by 24 GeV/c protons to $\Phi_{eq}$ between 0 and $1.3 \times 10^{16}$ cm$^{-2}$. At $t = 0$ a 50 ps 1 V ramp on the electrode at $y = 0$ is used to simulate the $y$ dependence of $E_W(y,t)$.

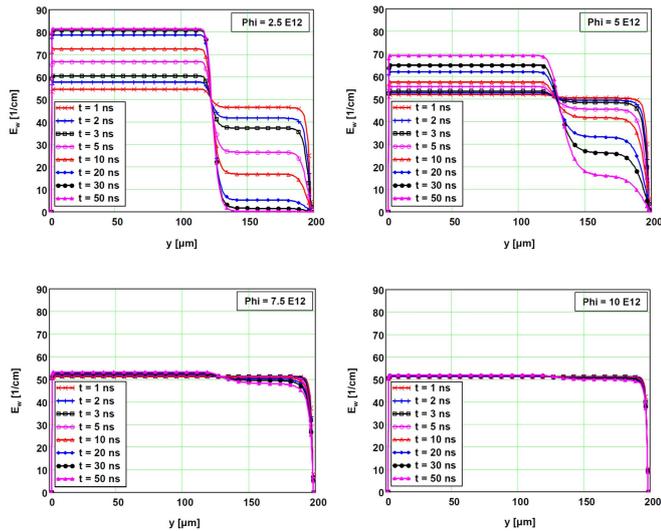

Fig. 4. y dependence of $E_W(y,t)$ for different times t of the pad diode at 40 V (w = 120 μm) for $\Phi_{eq}$ between 2.5 and $10 \times 10^{12}$ cm$^{-2}$.

Fig. 4 shows the results at 40 V for $\Phi_{eq}$ between 2.5 and $10 \times 10^{12}$ cm$^{-2}$. For $\Phi_{eq} \lesssim 2.5 \times 10^{12}$ cm$^{-2}$, $E_W(y,t)$ is the same as for the non-irradiated sensor: $1/d$ at $t = 0$, increasing to $1/w$ for $y < w$, and decreasing to 0 for $y > w$. For $\Phi_{eq} \gtrsim 10^{13}$ cm$^{-2}$, $E_W(y,t) = 1/d$, independent of $y$, $t$ and $V$.

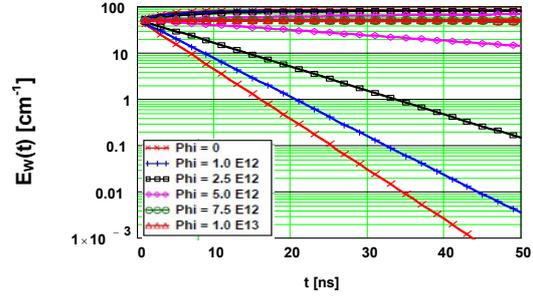

Fig. 5. Time dependence of $E_W$ for a sensor irradiated to different $\Phi_{eq}$ values. Curves with positive slopes are for $y = 20$ μm (depleted region), and with negative ones for $y = 180$ μm (non-depleted region). All curves are compatible with the exponential dependence of Eq. 1.

Fig. 5 shows the time dependence of $E_W$ in the depleted region (increasing curves), and in the non-depleted region (decreasing curves) for different $\Phi_{eq}$. The time dependence is exponential as expected from Eq. 1. The time constant increases from $\approx 2.5$ ns for $\Phi_{eq} < 10^{12}$ cm$^{-2}$ to $\approx 20$ μs. For $\Phi_{eq} \gtrsim 10^{13}$ cm$^{-2}$ the time constant is much longer than the integration time of the typical readout electronics. Thus $E_W$ can be considered time independent with the value $E_W = 1/d$, which is the weighting field of the fully depleted sensor before irradiation.

As a cross check of the method the relation $\tau = \varepsilon_{Si} \cdot \rho \cdot d/w$ is used to determine $\rho(\Phi_{eq})$ and compare it to $\rho$ extracted from the electron and hole densities and the electric field of the TCAD simulation. The result is shown in Fig. 6. The two values agree: $\rho$ increases from $\approx 2.5$ to $\approx 3 \times 10^4$ kΩ·cm. The first value corresponds to the B-doping density of the non-irradiated sensor, the latter to the intrinsic resistivity.

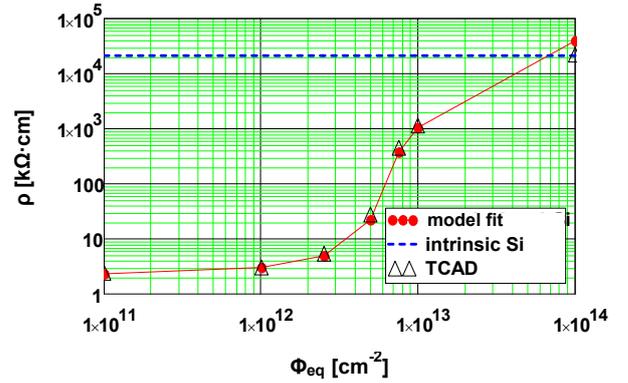

Fig. 6. $\Phi_{eq}$-dependence of $\rho$ in the low-field region at −30 °C from the TCAD simulation and from the model fit to $E_W(t)$.

IV. ADMITTANCE

The admittance $Y$, the inverse of the complex resistance $Z$, is closely related to the weighting field $E_W$, which describes the capacitive coupling of a charge in the sensor to the readout electrode. In the previous section it was shown that the time and position dependence of $E_W$ is caused by the resistivity $\rho$ of the low-field region. In this section the influence of $\rho$ on $Y$ is investigated.

The admittance $Y(f,U,\Phi_{eq})$ of 200 μm thick $n^+p$ pad sensors with $A = 0.25$ cm$^2$ area has been measured at –20 and –30 °C for frequencies $f$ between 100 Hz and 1 MHz and voltages $U$ between 1 and 1000 V. The sensors had been irradiated by 24 GeV/c protons to fluences $\Phi_{eq}$ of 3, 6, 8, and 13×10$^{15}$ cm$^{-2}$ and annealed for 80 min at 60 °C.

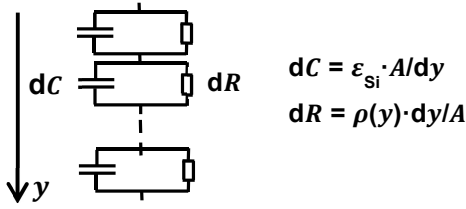

Fig. 7. Model for the frequency dependence of the admittance $Y$ of a pad sensors.

Fig. 7 shows the electrical model used to describe the data. The sensor is divided into slices of depth d$y$, each one described by a resistor d$R$ in parallel with a capacitor d$C$. The complex resistance $Z$ of the entire sensor is

$$Z(\omega) = \int_0^d \left(\frac{1}{dR} + i \cdot \omega \cdot dC\right)^{-1} = \frac{1}{A} \cdot \int_0^d \frac{\rho(y) \cdot dy}{1 + i \cdot \omega \cdot \rho(y)} \quad (2)$$

with $\omega = 2\pi f$ and $Y(\omega) = 1/Z(\omega)$. This model assumes that the $f$ dependence is caused only by the position-dependent resistivity $\rho(y)$, which is given by the density of electrons and holes and their respective mobilities. In the low-field region generation and recombination are in equilibrium, $n_e \cdot n_h = n_i^2$. The reduction of the generation-recombination lifetime with $\Phi_{eq}$ due to radiation-induced traps causes a decrease of the density of majority charge carriers and therefore an increase of $\rho$ until the intrinsic resistivity $\rho_{intr} = 1/(2 \cdot q_0 \cdot n_i \cdot \sqrt{\mu_e \cdot \mu_h})$ is reached [5]. The elementary charge is $q_0$, the intrinsic charge carrier density $n_i$, the electron and hole densities $n_e$ and $n_h$, and the electron and hole low-field mobilities $\mu_e$ and $\mu_h$. In the high-field region the free charge carriers drift in the electric field, and their density product is lower than $n_i^2$, and $\rho > \rho_{intr}$.

As $Y$ is insensitive to $y$, regions of equal $\rho$ are grouped and the variable $\eta$ is used instead of $y$. The resistivity is parametrized as $\rho(\eta) = \rho_0 + \rho_1 \cdot e^{-\eta/\lambda}$ with the free parameters $\rho_0$, $\rho_1$ and $\lambda$. For every voltage a combined $\chi^2$-fit of Im($Y/\omega$) to the measured parallel capacitance $C_p$, and of atan(Im($Y$)/Re($Y$)) to the measured phase $\varphi_Y$ is performed. Figs. 8 and 9 compare the fit results (lines) to the experimental data (symbols) for $\Phi_{eq}$ = 3×10$^{15}$ and 13×10$^{15}$ cm$^{-2}$ at –30 °C for selected voltages. It is found that the model describes the $f$ dependence of $C_p$ within ≈ 0.5 % and $\varphi_Y$ within ≈ 0.5°. Fits with 4 different $\rho(\eta)$ parametrizations have been performed. They describe the data too, and the results for $\rho(\eta)$, in particular in the low-field region and the transition to the high-field region, agree.

It is concluded that the $f$ dependence of the admittance $Y$ of a radiation-damaged Si pad sensor can be described by a model with a position-dependent resistivity and that it is not required to include the response times of the radiation-induced traps.

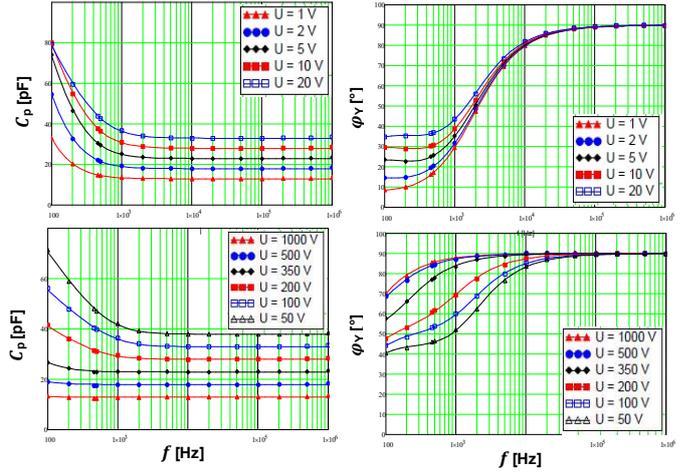

Fig. 8. Comparison of the measured parallel capacitance $C_p$ and of the admittance phase $\varphi_Y$ to the fit results for $\Phi_{eq} = 3\times10^{15}$ cm$^{-2}$ at –30 °C. The symbols are the experimental data and the solid lines the fit results. Top: $U$ = 1 to 20 V; bottom $U$ = 50 to 1000 V. For clarity the individual $C_p$ curves are shifted by 5 pF

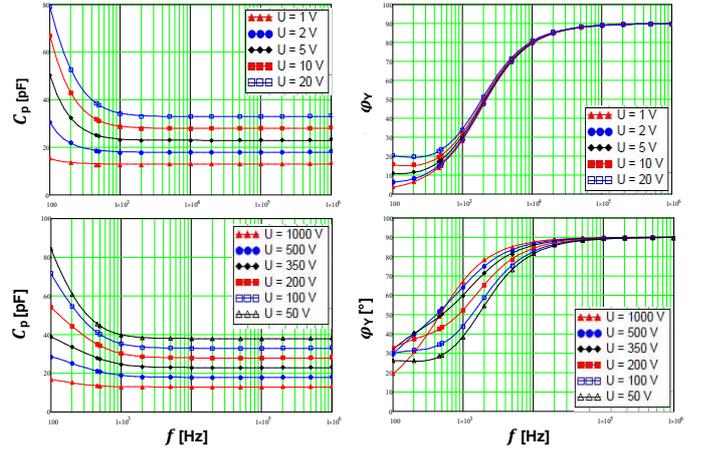

Fig. 9. Same as Fig. 8 for $\Phi_{eq} = 13\times10^{15}$ cm$^{-2}$.

Fig. 10 shows the distributions of $\rho(\eta)$ at –30 °C for the four $\Phi_{eq}$ values of the measurements. The low-field resistivities determined are $\rho_{low}$ = 81 MΩ·cm at –30 °C and 27 MΩ·cm at –20 °C; the corresponding values for $\rho_{intr}$ 70 and 27 MΩ·cm. Within the systematic uncertainties $\rho_{low}$ and $\rho_{intr}$ agree. The difference cannot be explained by an increase of $\rho_{low}$ with irradiation: The ratio $\rho_{low}/\rho_{intr}$ is independent of $\Phi_{eq}$.

Fig. 10 also shows that for a given $\Phi_{eq}$ the extension of the low-$\rho$ region decreases with voltage and increases with $\Phi_{eq}$. This agrees with the determination of the electric field $E$ in radiation-damaged strip detectors using the edge-TCT technique, where regions with low position-independent values of $E$ are observed [7, 8]. The values agree with the expectation $E = j \cdot \rho_{intr}$ from the ohmic voltage drop of the measured dark current density $j$ in a conductor with resistivity $\rho_{intr}$.

The region with $\rho \gg \rho_{intr}$ is the depletion region of the irradiated sensor. It is expected that when ionizing radiation is detected, most of the signal comes from this region. The increase of the high-$\rho$ region with $U$ and its decrease with $\Phi_{eq}$ qualitatively agrees with the observed charge collection efficiency. However, more work is needed to find out if and how $\rho(\eta)$ can be used to characterize radiation-damaged sensors.

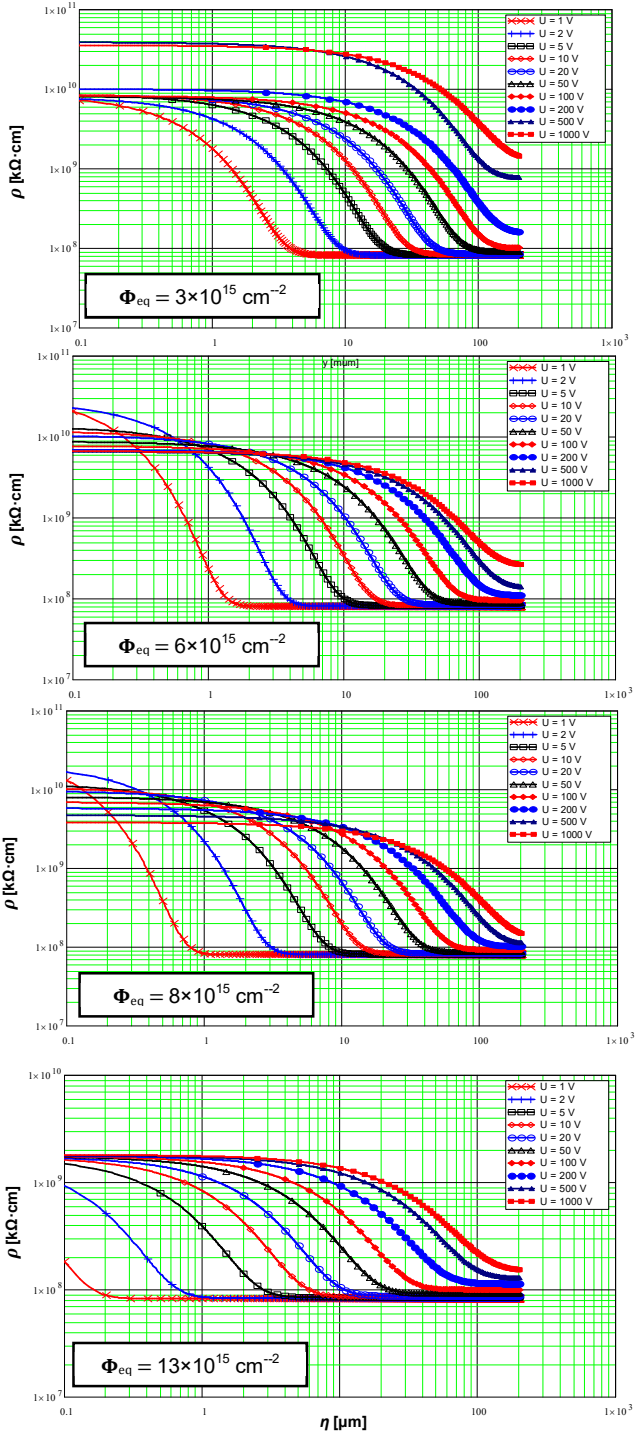

Fig. 10. $\rho(\eta)$ for selected voltages and different $\Phi_{eq}$ at –30 °C. For the definition of $\eta$ see text.

## V. SUMMARY AND CONCLUSIONS

In this paper the weighting field $E_W$ and the admittance $Y = 1/R_p + i \cdot \omega \cdot C_p$ of silicon pad sensors before and after hadron irradiation are studied; $R_p$ is the parallel resistance and $C_p$ the parallel capacitance. As $E_W$ describes the capacitive coupling of a charge in the sensor to the readout electrode, $C_p$ and $E_W$ are closely related.

For the study $n^+p$ silicon pad sensors fabricated on B-doped silicon with resistivity $\rho = 2.5$ kΩ·cm, thickness $d = 200$ μm, and area $A = 0.25$ cm$^2$ are used. The sensors were irradiated by 24 GeV/c protons to 1 MeV neutron equivalent fluences $\Phi_{eq} =$ 3 to 13×10$^{15}$ cm$^{-2}$ and annealed for 60 min at 80 °C.

First a partially depleted pad sensor with the depletion depth $w$ is considered. It is shown that $C_p$ depends on the frequency $f$: For $f \gg 1/\tau_{rel}$: $C_p = \varepsilon_{Si} \cdot A/d$, and for $f \ll 1/\tau_{rel}$: $C_p = \varepsilon_{Si} \cdot A/w$; with the dielectric relaxation time $\tau_{rel} = \varepsilon_{Si} \cdot \rho \approx 2.5$ ns. Also $E_W$ is found to depend on time: At $t = 0$, $E_W = 1/d$ in the entire sensor. As a function of $t$, $E_W$ approaches $1/w$ with the time constant $\tau = \tau_{rel} \cdot d/w$ in the depleted region. In the non-depleted region $E_W \to 0$ with the same $\tau$. Hadron irradiation increases $\rho$ from 2.5 kΩ·cm, given by the doping density, to the intrinsic resistivity, which has the value $\rho_{intr} \approx 70$ MΩ·cm at –30 °C. Simulations of irradiated sensors show that the increase of $\rho$ occurs for $\Phi_{eq}$ values between 10$^{12}$ and 10$^{14}$ cm$^{-2}$. The simulations also show that for the time dependence of $E_W$ for the irradiated sensor the same formula for the time constant $\tau$ can be used as for the non-irradiated sensor. The increase of $\rho$ with $\Phi_{eq}$ results in an increase of $\tau$ from a few ns to ≈ 100 μs. Given that the typical integration time of the electronics used for the readout of silicon sensors is tens of nanoseconds, a constant $E_W \approx 1/d$ can be assumed.

It is concluded that for calculating the signal in radiation-damaged silicon sensors the time-independent weighting field $E_W$ of the fully depleted sensor before irradiation can be used.

For the study of the frequency dependence of $C_p$ for radiation-damaged silicon sensors, the admittance $Y$ of pad sensors irradiated to $\Phi_{eq} =$ 3, 6, 8 and 13×10$^{15}$ cm$^{-2}$ has been measured for frequencies $f$ between 100 Hz and 1 MHz and reverse voltages $U = 1$ to 1000 V at –20 and –30 °C. The parallel capacitance $C_p$ and the phase of $Y$, $\varphi_Y$, for every value of $U$ has been fitted by a model with a position-dependent resistivity. The model provides a description of the data with rms deviations of ≈ 0.5 % for $C_p$, and ≈ 0.5° for $\varphi_Y$. Up to a certain voltage, which increases with $\Phi_{eq}$, regions with $\rho \approx \rho_{intr}$ are observed. In these regions the generation and recombination of electrons and holes are in equilibrium and the electric field is the result of the dark current in a conductor with the resistivity $\rho_{intr}$. The regions with $\rho \gg \rho_{intr}$ correspond to the depleted regions, from which most of the signal is expected, if the sensor is used for the detection of radiation.

It is concluded that the frequency dependence of irradiated sensors can be modelled with a position-dependent resistivity. The introduction of the frequency dependence of the radiation-induced traps is not required.